# Seeded growth of high-quality transition metal dichalcogenide single crystals via chemical vapor transport

*Hao Li, Junku Liu, Nan Guo, Lin Xiao, Haoxiong Zhang, Shuyun Zhou, Yang Wu,\* Shoushan Fan*

**Hao Li** — School of Materials Science and Engineering & Tsinghua-Foxconn Nanotechnology Research Center, Tsinghua University, Beijing, 100084, P. R. China

**Junku Liu** — Qian Xuesen Laboratory of Space Technology, China Academy of Space Technology, Beijing 100094, P. R. China

**Nan Guo** — Qian Xuesen Laboratory of Space Technology, China Academy of Space Technology, Beijing 100094, P. R. China

**Lin Xiao** — Qian Xuesen Laboratory of Space Technology, China Academy of Space Technology, Beijing 100094, P. R. China

**Haoxiong Zhang** — State Key Laboratory of Low Dimensional Quantum Physics and Department of Physics, Tsinghua University, Beijing 100084, P. R. China

**Shuyun Zhou** — State Key Laboratory of Low Dimensional Quantum Physics and Department of Physics, Tsinghua University, Beijing 100084, P. R. China; Frontier Science Center for Quantum Information, Beijing 100084, China

**Yang Wu** — Department of Mechanical Engineering & Tsinghua-Foxconn Nanotechnology Research Center, Tsinghua University, Beijing, 100084, P. R. China

**Shoushan Fan** — State Key Laboratory of Low Dimensional Quantum Physics and Department of Physics & Tsinghua-Foxconn Nanotechnology Research Center, Tsinghua University, Beijing, 100084, P. R. China

**ABSTRACT:** Transition metal dichalcogenides (TMDs) are van der Waals layered materials with sizable and tunable bandgaps, offering promising platforms for two-dimensional electronics and optoelectronics. To this end, the bottleneck is how to acquire high-quality single crystals in a facile and efficient manner. As one of the most widely employed method of single-crystal growth, conventional chemical vapor transport (CVT) generally encountered problems including the excess nucleation that leads to small crystal clusters and slow growth rate. To address these issues, a seed crystal is introduced to suppress the nucleation and an inner tube is adopted as both a separator and a flow restrictor, favoring the growth of large-size and high-quality TMD single crystals successfully. Three examples are presented, the effective growth of millimeter-sized $MoSe_2$ and $MoTe_2$ single crystals, and the greatly shortened growth period for $PtSe_2$ single crystal, all of which are synthesized in high quality according to detailed characterizations. The mechanism of seeded CVT is discussed. Furthermore, a phototransistor based on exfoliated multi-layered $MoSe_2$ displays excellent photoresponse in ambient conditions, and considerably rapid rise and fall time of 110 and 125 $\mu$s are obtained. This work paves the way for developing a facile and versatile method to synthesize high-quality TMD single crystals in laboratory, which could serve as favorable functional materials for potential low-dimensional optoelectronics.

**INTRODUCTION**

Transition metal dichalcogenides (TMDs) represent a category of materials with a general formula of $MCh_2$, where M is the central transition metal atom (group IV, V, VI, VII, IX or X) and Ch is the chalcogen atom (S, Se or Te). The typical crystal structure of TMDs can be regarded as van der Waals stacking of $MCh_2$ layers where the central M layer is sandwiched between two chalcogen layers, exhibiting a large variety of chemical composition and structural variations. Thus, TMDs have generated intriguing properties that triggered tremendous interests in both fundamental research and future technology.[1-11] For instance, $MoSe_2$ and $MoTe_2$ in 2H phase show a layered hexagonal structure, similar to graphene but featured an indirect-to-direct bandgap transition from bulk to monolayer in the range of 1-2 eV.[12,13] $PtSe_2$ in 1T phase experiences a semimetal to narrow bandgap semiconductor transition when it is thinned to bilayer and monolayer, suggesting a promising platform for the optoelectronics in the mid-infrared region.[14,15] With the bandgap tunability, TMDs outstrip gapless graphene by displaying high performance in electronics, optoelectronics and photovoltaics.[5,6] Furthermore, emerging topological states, such as Dirac semimetal and Weyl semimetal, have been realized in the 1T phase of $PtSe_2$,[16,17] $PtTe_2$,[18] $PdTe_2$[19,20] and $T_d$-$MoTe_2$,[21,22] respectively.

Noticeably, high-quality TMD single crystals are bestowed with advantages such as fewer defects over their film counterparts grown by chemical or physical vapor deposition, essential for discovering exotic physical phenomena including the nontrivial band structure and quantum transport.[8,23] Furthermore, the large-size and high-quality crystals are prerequisites for fabricating large-area few- or monolayer TMDs through mechanical or liquid exfoliation for low dimensional electronic and optoelectronic devices.[24-28] Among numerous single-crystal growth strategies, chemical vapor transport (CVT) has been extensively applied for scientific research. Generally, in CVT process, the condensed source is vaporized by a transport agent and recrystallizes at the deposition end, driven by the shift of chemical equilibrium in an established temperature gradient according to the Le Chatelier's principle.[29-35] However, the challenge in conventional CVT (C-CVT) is the excess nucleation that randomly takes place at multiple sites on the inner wall of the reaction vessel, leading to aggregated crystal clusters with restricted size. Therefore, it is difficult for C-CVT to synthesize large-size and high-quality TMD crystals that meet the demand in research and application. Adding a seed crystal at the deposition end could prevent such unwanted nucleation and significantly improve the crystal growth. To this far, seeded growth in CVT are mainly reported on the growth of ZnO and $Cu_2OSeO_3$ bulk crystal,[36-39] while its attempt in TMD crystal growth is yet to be explored.

In this study, we present a seeded CVT (S-CVT) method for the facile and effective growth of high-quality TMD crystals (cf. Figure 1). Compared to C-CVT, a seed crystal aiming to promote and control the growth of TMD crystals is introduced, and an inner tube serving as both a separator and a flow restrictor is added. S-CVT can effectively address the common issues such as random and excess nucleation and unsuitable growth substrate. As a result, millimeter-sized high-quality $MoSe_2$ and $MoTe_2$ single crystals with clearly hexagonal and shiny surface are successfully grown. Moreover, S-CVT can evidently shorten the growth period for strenuous $PtSe_2$ from 21 to 3 days, which is a type-II Dirac semimetal and a potential candidate for mid-infrared optoelectronics. The mechanism of S-CVT reaction is also discussed in detail. Furthermore, the highly crystalline nature of the as-grown TMD crystals are established through various characterizations and excellent photoresponse of a multi-layered $MoSe_2$ phototransistor in ambient environment.

**EXPERIMENTAL SECTION**

**Materials.** Mo foil (Alfa Aesar, 99.95%), Pt wire (Alfa Aesar, 99.9%), Se shot (Alfa Aesar, 99.999%), Te lump (Aladdin, 99.999%), $SeCl_4$ powder (Aladdin, 99.5%), $TeBr_4$ powder (Alfa Aesar, 99.9%).

**Synthesis of $MCh_2$ powder.** Polycrystalline $MCh_2$ (M = Mo, Pt, Ch = Se, Te) powder used as precursor was synthesized by direct solid-state reaction of the stoichiometric mixture of high-purity M and Ch. The mixture vacuum-sealed in a silica ampoule was heated to 1073 K and kept at the same temperature for 3 days.

**C-CVT Growth.** As shown in Figure S1, $MCh_2$ powder and transport agent were vacuum-sealed in a silica ampoule and the growth was carried out under the temperature gradient (see Figure S2) established in a horizontal tube furnace. The detailed parameters for C-CVT growth are listed in Table S1, which has been learned from previous literatures and optimized in plenty of experiments.[17,32,40] The as-grown crystals were rinsed with deionized $H_2O$, acetone and ethanol to remove remaining transport agent. Afterwards, shiny crystals with width of 500 μm - 1 mm and high crystallinity were carefully chosen and cut off into regular shape to be used as seed crystals for S- CVT growth (see Figure S3, S4).

**S-CVT Growth.** As shown in Figure 1, $MCh_2$ powder and transport agent were added into an inner silica tube horizontally placed at the source end of a vacuum-sealed silica ampoule, while the seed crystal was placed at the other end for the crystal growth. Both C-CVT and S-CVT were carried out in the same tube furnace and under the same condition. The detailed parameters for S-CVT growth are listed in Table S1.

**Characterization.** Powder X-ray diffraction (PXRD) patterns were collected on a Rigaku D/max-2500/PC X-ray diffractometer, using Cu Kα radiation and operating at 40 kV and 200 mA. Raman spectra were collected on a Horiba Jobin Yvon LabRam-HR/VV Spectrometer with a 514-nm laser source and a grating with 600 lines per millimeter. High-resolution transmission electron microscopy (HRTEM) images and selected area electron diffraction (SAED) patterns were collected on an FEI Tecnai G2F20 transmission electron microscope operating at 200 kV with an aperture size of 100 nm in diameter.

**Device Fabrication and measurement.** Back-gated multi-layered MoSe$_2$ phototransistor was fabricated in the following way. First, source, drain, and gate electrodes were patterned on a 300-nm SiO$_2$/p$^+$-Si substrate using standard UV photolithography techniques, followed by selective etching of 300-nm SiO$_2$ beneath the gate electrode and e-beam evaporation of a 5/100-nm Cr/Au film. Second, a multi-layered MoSe$_2$ sample obtained by mechanical exfoliation of an S-CVT grown MoSe$_2$ crystal was prepared on another 300-nm SiO$_2$/p$^+$-Si substrate. Finally, the sample was transferred onto patterned source-drain electrodes using polyvinyl alcohol (PVA) as a medium, which was removed in H$_2$O and rinsed with isopropyl alcohol. Electrical measurements were carried out in dark and ambient environments. Electrical and photoresponse measurements were conducted on a Lake Shore Probe Station with an Agilent B1500 semiconductor parameter analyzer and a laser diode with a wavelength of 637 nm. The device was illuminated with a laser spot size larger than 200 μm to ensure a uniform intensity. The modulation cycle of temporal response was recorded using a current preamplifier and an oscilloscope.

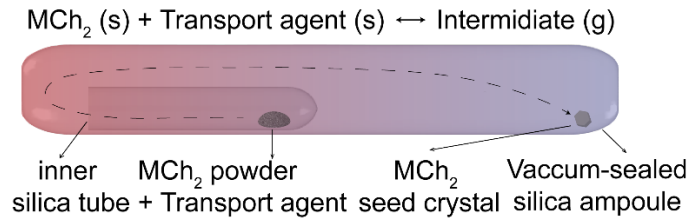

**Figure 1.** Schematic diagram of the S-CVT growth setup.

## RESULTS AND DISCUSSION

As shown in Figure 2, it can be learned from plenty of experiments that C-CVT generally produces MoSe$_2$ and MoTe$_2$ crystals with limited width of 500 μm - 1 mm that congregate together (see Figure 2d, e). In contrast, S-CVT can produce large-size MoSe$_2$ and MoTe$_2$ crystals up to 3 mm with clearly hexagonal shape and flat surface (see Figure 2a, b). For PtSe$_2$ grown by C-CVT, it requires a long growth period of at least 3 weeks to provide samples with satisfactory quality.[17] In comparison, S-CVT can evidently reduce the growth period to 3 days and increase the crystal size in the meantime. Both PtSe$_2$ crystals grown by S-CVT and C-CVT are shown in Figure 2c and f, displaying hexagonal shape and shiny surface. The observable boundaries are possibly arising from stress and cracks in CVT systems.

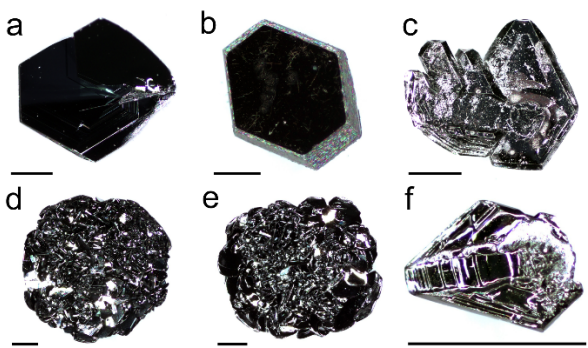

**Figure 2.** Optical images of (a) MoSe$_2$, (b) MoTe$_2$, (c) PtSe$_2$ crystals grown via S-CVT and (d) MoSe$_2$, (e) MoTe$_2$, (f) PtSe$_2$ crystals grown via C-CVT. The scale bars are 1 mm.

Figure 3a and b display PXRD patterns of the raw MoSe$_2$ and MoTe$_2$ crystals grown via C-CVT and S-CVT, respectively. The former patterns show no obvious peaks, while the later patterns contain sharp and intense peaks that follow the (0 0 l), l = 2n, diffraction rule, indicating aligned MCh$_2$ layers along the [0 0 1] direction and high degree of crystallization of MoSe$_2$ and MoTe$_2$ crystals grown via S-CVT. Figure 3c displays PXRD patterns of slightly ground PtSe$_2$ crystals grown via C-CVT and S-CVT, respectively. The sharp and intense peaks are in good agreement with the $P\bar{3}m1$ PtSe$_2$ (PDF#18-0970), revealing the good crystallinity. Raman spectra (Figure 4a-c) show the characteristic vibrational modes of MoSe$_2$, MoTe$_2$ and PtSe$_2$, consistent with the literature.[12,17,41] No shift is observed between crystals grown via C-CVT and S-CVT. Figure 5a-c present HRTEM images of MoSe$_2$, MoTe$_2$ and PtSe$_2$ crystals grown via S-CVT. The lattice fringes are clearly resolved and spacings of (1 0 0) planes match well with the crystallographic values of 0.28, 0.30 and 0.32 nm in MoSe$_2$, MoTe$_2$, and PtSe$_2$, respectively. In addition, the highly crystalline nature is again reflected through luminous dots in the SAED patterns (insets of Figure 5a-c), where all samples show one uniform diffraction pattern with hexagonal symmetry contributed by well-crystallized MCh$_2$ layers.

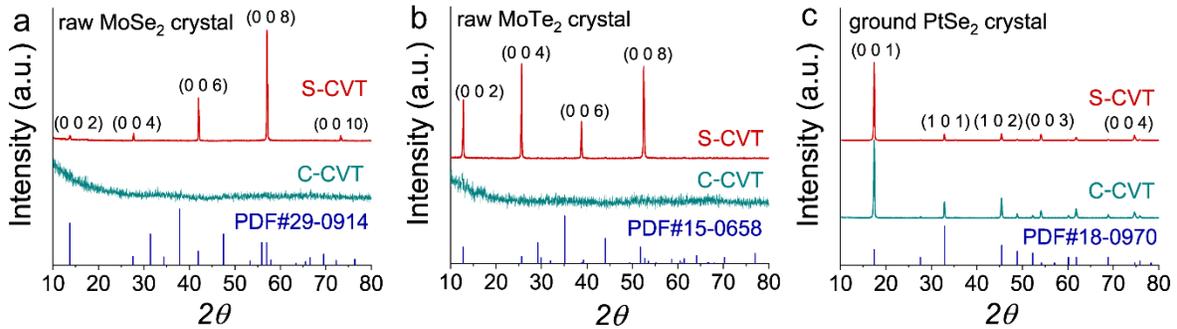

**Figure 3.** PXRD patterns of (a) MoSe$_2$, (b) MoTe$_2$ crystals, and (c) ground PtSe$_2$ crystals grown via S-CVT and C-CVT.

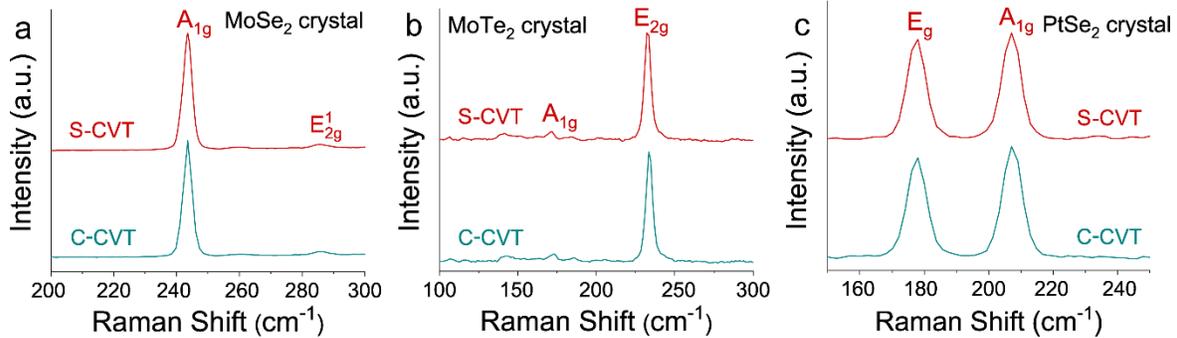

**Figure 4.** Raman spectra of (a) MoSe$_2$, (b) MoTe$_2$, (c) PtSe$_2$ crystals grown via S-CVT and C-CVT.

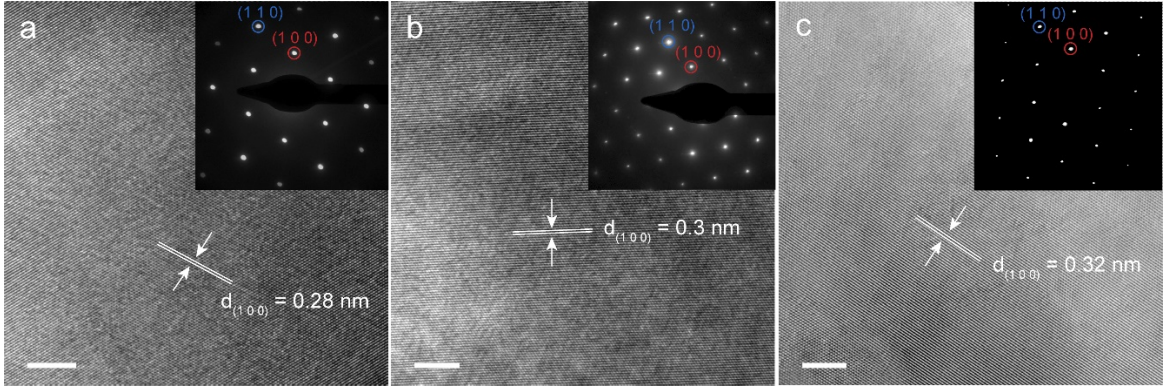

**Figure 5.** HRTEM images of (a) MoSe$_2$, (b) MoTe$_2$, (c) PtSe$_2$ crystals grown via S-CVT. The scale bars are 5 nm. Insets are SAED patterns

The S-CVT affords more controllable manners to produce high-quality MCh$_2$ crystals. Taking into account the growth of MoSe$_2$ crystal as an example, SeCl$_4$ is used as a transport agent to generate Cl$_2$ that transfers MoSe$_2$ during the CVT. The transport process can be described according to the following equilibria:

$$2 \text{ SeCl}_4 \text{ (s)} \leftrightharpoons \text{Se}_2 \text{ (g)} + 4 \text{ Cl}_2 \text{ (g)} \qquad (1)$$

$$2 \text{ MoSe}_2 \text{ (s)} + x \text{ Cl}_2 \text{ (g)} \leftrightharpoons 2 \text{ MoCl}_x \text{ (g)} + 2 \text{ Se}_2 \text{ (g)} \qquad (2)$$

where MCl$_x$ represents gaseous species of the metal chloride (MCl$_2$, MCl$_3$, MCl$_4$, MCl$_5$, ...).[31,32] At the source end, the equilibrium (2) proceeds forward to generate gaseous intermediate MoCl$_x$. Due to the temperature gradient in the reaction ampoule (see Figure S2), a partial pressure gradient of MoCl$_x$ is established, which pushes the equilibrium (2) backward at the deposition end. For C-CVT, the nucleation of MoSe$_2$ randomly occurs at multiple sites on the wall of silica ampoule with no preferred orientation, producing the congregated small MoSe$_2$ crystals (see Figure 2d). For S-CVT, the MoSe$_2$ seed crystal favors the nucleation rather than the silica wall due to the lower nucleation energy, which in turn consumes the gaseous intermediate MoCl$_x$ at the deposition end and impedes extra nucleation. In the meantime, the MoSe$_2$ seed crystal (see Figure S3a, S4a) encourages oriented nucleation that favors the growth of high-quality single crystals. Moreover, an inner silica tube is added to separate the seed crystal and source material. The inner silica tube also acts as a flow restrictor, where the slightly higher temperature of ~10 K at the opening end (see Figure S2) slows down the transport of gaseous MoCl$_x$. Consequently, it can decrease the quantity of nucleation centers on the surface of seed crystal, reducing defects and promoting the quality of as-grown single crystals. Similar mechanism also works for the growth of MoTe$_2$ crystal. However, for the growth of PtSe$_2$ crystal in C-CVT method, there is a major difference that its growth is arduous and sluggish compared to those of MoSe$_2$ and MoTe$_2$, where the nucleation plays a more important role. It typically requires long growth periods of no less than 3 weeks in C-CVT method, while the seed crystal can relieve this issue and remarkably promote the growth of PtSe$_2$ by reducing its growth period to just 3 days in S-CVT system.

Owing to the unique electronic structures and sizable bandgaps, TMD materials could serve as promising functional materials for future low dimensional electronics and optoelectronics.[6-8] On account of this, we fabricated back-gated multi-layered MoSe$_2$ phototransistor based on the mechanical exfoliation of MoSe$_2$ crystal grown via S-CVT (see inset of Figure 6a). A linear dependence of the output characteristics (see Figure S5a) implies the ohmic character of contacts between the MoSe$_2$ flake and electrodes. The transfer characteristics in Figure S5b reveal an air-stable n-type behavior, where the on-state is at positive gate voltage and the off-state is at negative gate voltage. The relatively large threshold voltage of -30 V and low field-effect mobility of 0.77 cm$^2$ V$^{-1}$ s$^{-1}$ might ascribe to the defects and disorders introduced by mechanical exfoliation, device fabrication and SiO$_2$ substrate.[7,42-45]

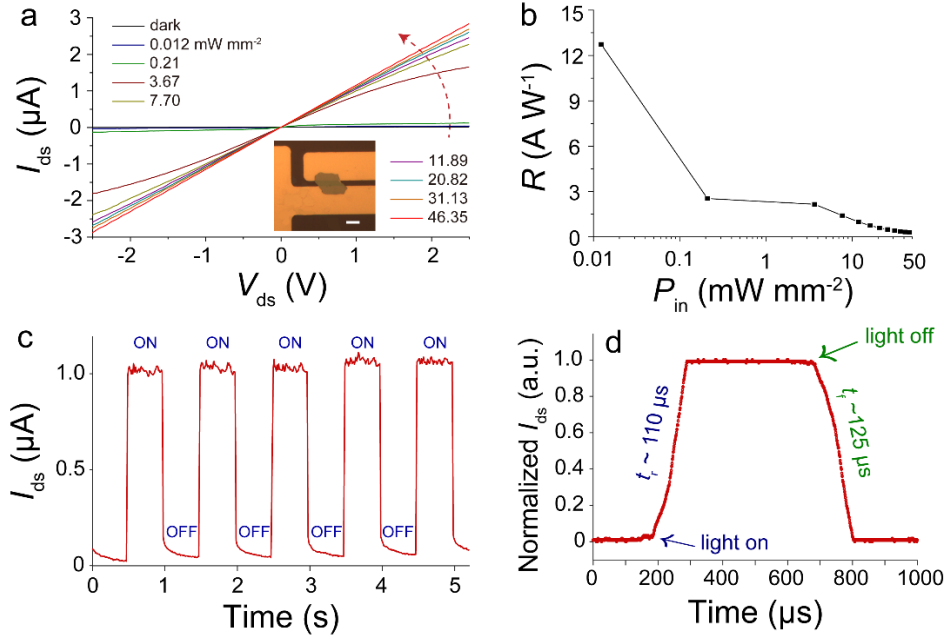

**Figure 6.** Photoresponse performance of the back-gated multi-layered MoSe$_2$ phototransistor under 637-nm illumination. (a) The drain-source current ($I_{ds}$) versus the drain-source voltage ($V_{ds}$) curves with different light power intensities ($P_{in}$) at $V_{bg}$ = -50 V. The arrow direction indicates increasing $P_{in}$. The inset in (a) is the optical image of the phototransistor, and the scale bar is 10 μm. (b) Photoresponsivity ($R$) as a function of $P_{in}$ at $V_{ds}$ = 2.5 V and $V_{bg}$ = -50 V. (c) Temporal response at $V_{ds}$ = 1 V, $V_{bg}$ = -50 V and $P_{in}$ = 20.82 mW mm$^{-2}$. The light is turned on and off at an interval of 1 s. (d) The rise and fall time are 110 and 125 $\mu$s, respectively, in a single modulation cycle.

The photoresponse performance of the multi-layered MoSe$_2$ phototransistor was measured under a 637-nm illumination in ambient condition. The curves of drain-source current ($I_{ds}$) versus voltage ($V_{ds}$) with different light power intensities ($P_{in}$) at $V_{bg}$ = -50 V are displayed in Figure 6a. The illumination causes a distinct increase in $I_{ds}$ and a net photocurrent ($I_{ph} = I_{illuminated} - I_{dark}$) as high as 2.8 $\mu$A is obtained at $V_{ds}$ = 2.5 V and $P_{in}$ = 46.35 mW mm$^{-2}$. At a fixed bias voltage, the increase of $P_{in}$ could induce the increase of $I_{ph}$, arising from the increase of photogenerated carriers. Figure 6b depicts the photoresponsivity ($R = I_{ph} / P_{in}S$, where $S$ is the area of the MoSe$_2$ flake) as a function of $P_{in}$ at $V_{ds}$ = 2.5 V and $V_{bg}$ = -50 V, and the maximum and minimum photoresponsivity are 13 and 0.3 A W$^{-1}$ at light power intensity of 0.012 and 46.35 mW mm$^{-2}$, respectively. For comparison, it has been reported that a typical back-gated phototransistor based on multi-layered MoSe$_2$ flake obtained by mechanical exfoliation exhibits a maximum photoresponsivity of ~1.4 A W$^{-1}$ at a light power of ~100 nW under 532-nm illumination in the off-state, while a top-gated photodetector based on monolayer MoSe$_2$ grown via CVD shows a lower photoresponsivity of ~13 mA W$^{-1}$ at $P_{in}$ = 1 mW mm$^{-2}$ under 532-nm illumination in the off-state.[46,47] Therefore, the MoSe$_2$ phototransistor in this work displays a high and stable photoresponsivity with a wide range of $P_{in}$. As shown in Figure 6b and S6, the photocurrent tends to reach saturation and the photoresponsivity decreases with the increasing $P_{in}$, mainly owing to the carrier-trap saturation effect resulting from surface state filling.[48-50] The time-resolved photoresponse was investigated to further evaluate the device performance. Figure 6c depicts the time-resolved drain current by switching the illumination light on and off periodically at $V_{ds}$ = 1 V, $V_{bg}$ = -50 V and $P_{in}$ = 20.82 mW mm$^{-2}$. The on-state and off-state are realized under illuminated and dark conditions, respectively, and the stable and reversible switching between them exhibits a highly robust feature of the multi-layered MoSe$_2$ phototransistor. Moreover, a single modulation cycle of temporal response is shown in Figure 6d. The current displays sharp rising and falling edges, with the corresponding rise and fall time of 110 and 125 μs, respectively. In contrast, the early reported multi-layered MoSe$_2$ phototransistor and the CVD-grown monolayer MoSe$_2$ photodetector display rise and fall time of 15 and 30 ms, 60 and 60 ms, respectively, demonstrating the considerably rapid photoresponse of the back-gated multi-layered MoSe$_2$ phototransistor in this work.[46,47] The rapid photoresponse can be attributed to the fast carrier recombination in the multi-layered MoSe$_2$, which is probably associated with the highly crystalline character of MoSe$_2$ crystals grown via S-CVT. It is noteworthy that the performance of the device could be further improved by optimizing the sample treatment and device fabrication.[7,24] The excellent photoresponse performance suggests that TMD crystals grown via S-CVT could serve as excellent materials for the construction of high-performance low dimensional optoelectronics.

**CONCLUSIONS**

In summary, S-CVT promotes the crystal growth through a seed crystal acting as nucleation and crystallization substrate. The inner tube in the growth apparatus acts as both a separator and a flow restrictor, on account of the reaction equilibrium and mass transport in the CVT system. It has been proved that S-CVT can be facile, sound, and versatile for large-size TMD crystal growth by laboratory-based apparatus, as exemplified in MoSe$_2$, MoTe$_2$, and PtSe$_2$. The high crystallinity of the as-grown TMD crystals has been well characterized via PXRD, Raman, HRTEM, and SAED. The excellent photoresponse performance of MoSe$_2$ grown by S-CVT that outperforms the exfoliated or CVD-grown ones suggests its promise for investigating emerging low-dimensional optoelectronics. Furthermore, considering the generality of its principle and versatility of the growth substrate, S-CVT could be extended to synthesize heterogeneous TMDs, for instance, the van der Waals heterojunctions, and be expected to become a powerful strategy for investigating emerging properties of 2D materials.

**AUTHOR INFORMATION**

**Corresponding Author**

**Yang Wu**, Email: wuyangthu@tsinghua.edu.cn

**Author Contributions**

The manuscript was written through contributions of all authors. All authors have given approval to the final version of the manuscript.

**Funding Sources**

This work was supported by the National Natural Science Foundation of China (Grant No. 21975140, 51991313, and 61904203).

**Notes**

The authors declare no competing financial interest.

**REFERENCES**

(1) Zhang, Y.; Yao, Y. Y.; Sendeku, M. G.; Yin, L.; Zhan, X. Y.; Wang, F.; Wang, Z. X.; He, J. Recent Progress in CVD Growth of 2D Transition Metal Dichalcogenides and Related Heterostructures. *Adv. Mater.* **2019,** 31, 1901694.

(2) Yun, Q. B.; Li, L. X.; Hu, Z. N.; Lu, Q. P.; Chen, B.; Zhang, H. Layered Transition Metal Dichalcogenide-Based Nanomaterials for Electrochemical Energy Storage. *Adv. Mater.* **2020,** 32, 1903826.

(3) Taghinejad, H.; Eftekhar, A. A.; Adibi, A. Lateral and vertical heterostructures in two-dimensional transition-metal dichalcogenides Invited. *Opt. Mater. Express* **2019,** 9, 1590-1607.

(4) Pi, L. J.; Li, L.; Liu, K. L.; Zhang, Q. F.; Li, H. Q.; Zhai, T. Y. Recent Progress on 2D Noble-Transition-Metal Dichalcogenides. *Adv. Funct. Mater.* **2019,** 29, 1904932.

(5) Zhu, W. J.; Low, T.; Wang, H.; Ye, P. D.; Duan, X. F. Nanoscale electronic devices based on transition metal dichalcogenides. *2D Mater.* **2019,** 6, 032004.

(6) Choi, W.; Choudhary, N.; Han, G. H.; Park, J.; Akinwande, D.; Lee, Y. H. Recent development of two-dimensional transition metal dichalcogenides and their applications. *Mater. Today* **2017,** 20, 116-130.

(7) Hu, Z.; Wu, Z.; Han, C.; He, J.; Ni, Z.; Chen, W. Two-dimensional transition metal dichalcogenides: interface and defect engineering. *Chem. Soc. Rev.* **2018,** 47, 3100-3128.

(8) Manzeli, S.; Ovchinnikov, D.; Pasquier, D.; Yazyev, O. V.; Kis, A. 2D transition metal dichalcogenides. *Nat. Rev. Mater.* **2017,** 2, 17033.

(9) Lin, L. X.; Lei, W.; Zhang, S. W.; Liu, Y. Q.; Wallace, G. G.; Chen, J. Two-dimensional transition metal dichalcogenides in supercapacitors and secondary batteries. *Energy Storage Mater.* **2019,** 19, 408-423.

(10) Li, Y. Z.; Shi, J.; Mi, Y.; Sui, X. Y.; Xu, H. Y.; Liu, X. F. Ultrafast carrier dynamics in two-dimensional transition metal dichalcogenides. *J. Mater. Chem. C* **2019,** 7, 4304-4319.

(11) Hu, H. W.; Zavabeti, A.; Quan, H. Y.; Zhu, W. Q.; Wei, H. Y.; Chen, D. C.; Ou, J. Z. Recent advances in two-dimensional transition metal dichalcogenides for biological sensing. *Biosens. Bioelectron.* **2019,** 142, 111573.

(12) Ruppert, C.; Aslan, O. B.; Heinz, T. F. Optical properties and band gap of single- and few-layer MoTe$_2$ crystals. *Nano Lett.* **2014,** 14, 6231-6236.

(13) Zhang, Y.; Chang, T.-R.; Zhou, B.; Cui, Y.-T.; Yan, H.; Liu, Z.; Schmitt, F.; Lee, J.; Moore, R.; Chen, Y.; Lin, H.; Jeng, H.-T.; Mo, S.-K.; Hussain, Z.; Bansil, A.; Shen, Z.-X. Direct observation of the transition from indirect to direct bandgap in atomically thin epitaxial MoSe$_2$. *Nat. Nanotechnol.* **2014,** 9, 111-115.

(14) Yan, M. Z.; Wang, E. Y.; Zhou, X.; Zhang, G. Q.; Zhang, H. Y.; Zhang, K. N.; Yao, W.; Lu, N. P.; Yang, S. Z.; Wu, S. L.; Yoshikawa, T.; Miyamoto, K.; Okuda, T.; Wu, Y.; Yu, P.; Duan,

W. H.; Zhou, S. Y. High quality atomically thin PtSe$_2$ films grown by molecular beam epitaxy. *2D Mater*. **2017,** 4, 045015.

(15) Yu, X.; Yu, P.; Wu, D.; Singh, B.; Zeng, Q.; Lin, H.; Zhou, W.; Lin, J.; Suenaga, K.; Liu, Z.; Wang, Q. J. Atomically thin noble metal dichalcogenide: a broadband mid-infrared semiconductor. *Nat. Commun*. **2018,** 9, 1545.

(16) Huang, H. Q.; Zhou, S. Y.; Duan, W. H. Type-II Dirac fermions in the PtSe$_2$ class of transition metal dichalcogenides. *Phys. Rev. B* **2016,** 94, 121117.

(17) Zhang, K. N.; Yan, M. Z.; Zhang, H. X.; Huang, H. Q.; Arita, M.; Sun, Z.; Duan, W. H.; Wu, Y.; Zhou, S. Y. Experimental evidence for type-II Dirac semimetal in PtSe$_2$. *Phys. Rev. B* **2017,** 96, 125102.

(18) Yan, M. Z.; Huang, H. Q.; Zhang, K. N.; Wang, E. Y.; Yao, W.; Deng, K.; Wan, G. L.; Zhang, H. Y.; Arita, M.; Yang, H. T.; Sun, Z.; Yao, H.; Wu, Y.; Fan, S. S.; Duan, W. H.; Zhou, S. Y. Lorentz-violating type-II Dirac fermions in transition metal dichalcogenide PtTe$_2$. *Nat. Commun*. **2017,** 8, 257.

(19) Fei, F. C.; Bo, X. Y.; Wang, R.; Wu, B.; Jiang, J.; Fu, D. Z.; Gao, M.; Zheng, H.; Chen, Y. L.; Wang, X. F.; Bu, H. J.; Song, F. Q.; Wan, X. G.; Wang, B. G.; Wang, G. H. Nontrivial Berry phase and type-II Dirac transport in the layered material PdTe$_2$. *Phys. Rev. B* **2017,** 96, 041201.

(20) Noh, H. J.; Jeong, J.; Cho, E. J.; Kim, K.; Min, B. I.; Park, B. G. Experimental Realization of Type-II Dirac Fermions in a PdTe$_2$ Superconductor. *Phys. Rev. Lett*. **2017,** 119, 016401.

(21) Deng, K.; Wan, G.; Deng, P.; Zhang, K.; Ding, S.; Wang, E.; Yan, M.; Huang, H.; Zhang, H.; Xu, Z.; Denlinger, J.; Fedorov, A.; Yang, H.; Duan, W.; Yao, H.; Wu, Y.; Fan, S.; Zhang, H.; Chen, X.; Zhou, S. Experimental observation of topological Fermi arcs in type-II Weyl semimetal MoTe$_2$. *Nat. Phys*. **2016,** 12, 1105-1110.

(22) Huang, L.; McCormick, T. M.; Ochi, M.; Zhao, Z. Y.; Suzuki, M. T.; Arita, R.; Wu, Y.; Mou, D. X.; Cao, H. B.; Yan, J. Q.; Trivedi, N.; Kaminski, A. Spectroscopic evidence for a type II Weyl semimetallic state in MoTe$_2$. *Nat. Mater.* **2016,** 15, 1155-1160.

(23) Armitage, N. P.; Mele, E. J.; Vishwanath, A. Weyl and Dirac semimetals in three-dimensional solids. *Rev. Mod. Phys.* **2018,** 90, 015001.

(24) Koppens, F. H.; Mueller, T.; Avouris, P.; Ferrari, A. C.; Vitiello, M. S.; Polini, M. Photodetectors based on graphene, other two-dimensional materials and hybrid systems. *Nat. Nanotechnol.* **2014,** 9, 780-793.

(25) Peng, J.; Wu, J.; Li, X.; Zhou, Y.; Yu, Z.; Guo, Y.; Wu, J.; Lin, Y.; Li, Z.; Wu, X.; Wu, C.; Xie, Y. Very Large-Sized Transition Metal Dichalcogenides Monolayers from Fast Exfoliation by Manual Shaking. *J. Am. Chem. Soc.* **2017,** 139, 9019-9025.

(26) Huang, Y.; Sutter, E.; Shi, N. N.; Zheng, J.; Yang, T.; Englund, D.; Gao, H.-J.; Sutter, P. Reliable Exfoliation of Large-Area High-Quality Flakes of Graphene and Other Two-Dimensional Materials. *ACS Nano* **2015,** 9, 10612-10620.

(27) Desai, S. B.; Madhvapathy, S. R.; Amani, M.; Kiriya, D.; Hettick, M.; Tosun, M.; Zhou, Y.; Dubey, M.; Ager, J. W., III; Chrzan, D.; Javey, A. Gold-Mediated Exfoliation of Ultralarge Optoelectronically-Perfect Monolayers. *Adv. Mater.* **2016,** 28, 4053-4058.

(28) Velicky, M.; Donnelly, G. E.; Hendren, W. R.; McFarland, S.; Scullion, D.; DeBenedetti, W. J. I.; Correa, G. C.; Han, Y.; Wain, A. J.; Hines, M. A.; Muller, D. A.; Novoselov, K. S.; Abruna, H. D.; Bowman, R. M.; Santos, E. J. G.; Huang, F. Mechanism of Gold-Assisted Exfoliation of Centimeter-Sized Transition-Metal Dichalcogenide Monolayers. *ACS Nano* **2018,** 12, 10463-10472.

(29) Binnewies, M.; Glaum, R.; Schmidt, M.; Schmidt, P. Chemical Vapor Transport Reactions - A Historical Review. *Z. Anorg. Allg. Chem.* **2013,** 639, 219-229.

(30) Wang, D.; Luo, F.; Lu, M.; Xie, X.; Huang, L.; Huang, W. Chemical Vapor Transport Reactions for Synthesizing Layered Materials and Their 2D Counterparts. *Small* **2019,** 15, e1804404.

(31) Ubaldini, A.; Jacimovic, J.; Ubrig, N.; Giannini, E. Chloride-Driven Chemical Vapor Transport Method for Crystal Growth of Transition Metal Dichalcogenides. *Cryst. Growth Des.* **2013,** 13, 4453-4459.

(32) Binnewies, M.; Glaum, R.; Schmidt, M.; Schmidt, P. *Chemical Vapor Transport Reactions*; Walter De Gruyter Gmbh: Berlin, 2012.

(33) Lenz, M.; Gruehn, R. Developments in measuring and calculating chemical vapor transport phenomena demonstrated on Cr, Mo, W, and their compounds. *Chem. Rev.* **1997,** 97, 2967-2994.

(34) Hu, D. K.; Xu, G. C.; Xing, L.; Yan, X. X.; Wang, J. Y.; Zheng, J. Y.; Lu, Z. X.; Wang, P.; Pan, X. Q.; Jiao, L. Y. Two-Dimensional Semiconductors Grown by Chemical Vapor Transport. *Angew. Chem. Int. Edit.* **2017,** 56, 3611-3615.

(35) Gruehn, R.; Glaum, R. New results of chemical transport as a method for the preparation and thermochemical investigation of solids. *Angew. Chem. Int. Edit.* **2000,** 39, 692-716.

(36) Skupiński, P.; Grasza, K.; Mycielski, A.; Paszkowicz, W.; Łusakowska, E.; Tymicki, E.; Jakieła, R.; Witkowski, B. Seeded growth of bulk ZnO by chemical vapor transport. *Phys. Status Solidi B* **2010,** 247, 1457-1459.

(37) Hong, S.-H.; Mikami, M.; Mimura, K.; Uchikoshi, M.; Yasuo, A.; Abe, S.; Masumoto, K.; Isshiki, M. Growth of high-quality ZnO single crystals by seeded CVT using the newly designed ampoule. *J. Cryst. Growth* **2009,** 311, 3609-3612.

(38) Fan, L.; Xiao, T.; Zhong, C.; Wang, J.; Chen, J.; Wang, X.; Peng, L.; Wu, W. Seeded growth of bulk ZnO crystals in a horizontal tubular furnace. *CrystEngComm* **2019,** 21, 1288-1292.

(39) Panella, J. R.; Trump, B. A.; Marcus, G. G.; McQueen, T. M. Seeded Chemical Vapor Transport Growth of $Cu_2OSeO_3$. *Cryst. Growth Des.* **2017,** 17, 4944-4948.

(40) Zhang, H.; Bao, C.; Jiang, Z.; Zhang, K.; Li, H.; Chen, C.; Avila, J.; Wu, Y.; Duan, W.; Asensio, M. C.; Zhou, S. Resolving Deep Quantum-Well States in Atomically Thin 2H-MoTe2 Flakes by Nanospot Angle-Resolved Photoemission Spectroscopy. *Nano Lett.* **2018,** 18, 4664-4668.

(41) Tonndorf, P.; Schmidt, R.; Bottger, P.; Zhang, X.; Borner, J.; Liebig, A.; Albrecht, M.; Kloc, C.; Gordan, O.; Zahn, D. R. T.; de Vasconcellos, S. M.; Bratschitsch, R. Photoluminescence emission and Raman response of monolayer $MoS_2$, $MoSe_2$, and $WSe_2$. *Opt. Express* **2013,** 21, 4908-4916.

(42) Liu, Y.; Guo, J.; Zhu, E.; Liao, L.; Lee, S. J.; Ding, M.; Shakir, I.; Gambin, V.; Huang, Y.; Duan, X. Approaching the Schottky-Mott limit in van der Waals metal-semiconductor junctions. *Nature* **2018,** 557, 696-700.

(43) Cui, X.; Shih, E. M.; Jauregui, L. A.; Chae, S. H.; Kim, Y. D.; Li, B.; Seo, D.; Pistunova, K.; Yin, J.; Park, J. H.; Choi, H. J.; Lee, Y. H.; Watanabe, K.; Taniguchi, T.; Kim, P.; Dean, C. R.; Hone, J. C. Low-Temperature Ohmic Contact to Monolayer $MoS_2$ by van der Waals Bonded Co/h-BN Electrodes. *Nano Lett.* **2017,** 17, 4781-4786.

(44) Liu, Y.; Guo, J.; Wu, Y.; Zhu, E.; Weiss, N. O.; He, Q.; Wu, H.; Cheng, H. C.; Xu, Y.; Shakir, I.; Huang, Y.; Duan, X. Pushing the Performance Limit of Sub-100 nm Molybdenum Disulfide Transistors. *Nano Lett.* **2016,** 16, 6337-6342.

(45) Bao, W.; Cai, X.; Kim, D.; Sridhara, K.; Fuhrer, M. S. High mobility ambipolar $MoS_2$ field-effect transistors: Substrate and dielectric effects. *Appl. Phys. Lett.* **2013,** 102, 042104.

(46) Xia, J.; Huang, X.; Liu, L. Z.; Wang, M.; Wang, L.; Huang, B.; Zhu, D. D.; Li, J. J.; Gu, C. Z.; Meng, X. M. CVD synthesis of large-area, highly crystalline $MoSe_2$ atomic layers on diverse substrates and application to photodetectors. *Nanoscale* **2014,** 6, 8949-8955.

(47) Abderrahmane, A.; Ko, P. J.; Thu, T. V.; Ishizawa, S.; Takamura, T.; Sandhu, A. High photosensitivity few-layered $MoSe_2$ back-gated field-effect phototransistors. *Nanotechnology* **2014,** 25, 365202.

(48) Guo, N.; Gong, F.; Liu, J.; Jia, Y.; Zhao, S.; Liao, L.; Su, M.; Fan, Z.; Chen, X.; Lu, W.; Xiao, L.; Hu, W. Hybrid $WSe_2$-$In_2O_3$ Phototransistor with Ultrahigh Detectivity by Efficient Suppression of Dark Currents. *ACS Appl. Mater. Interfaces* **2017,** 9, 34489-34496.

(49) Wu, P.; Dai, Y.; Ye, Y.; Yin, Y.; Dai, L. Fast-speed and high-gain photodetectors of individual single crystalline $Zn_3P_2$ nanowires. *J. Mater. Chem.* **2011,** 21, 2563-2567.

(50) Ahn, Y. H.; Park, J. Efficient visible light detection using individual germanium nanowire field effect transistors. *Appl. Phys. Lett.* **2007,** 91, 162102.